\documentclass[aps,pre,10pt,twocolumn,superscriptaddress]{revtex4-2}
\usepackage{graphicx}
\usepackage{epstopdf}
\usepackage{xcolor}
\usepackage{amsmath}
\usepackage{amssymb}
\usepackage[utf8]{inputenc}
\usepackage{graphicx,float,hyperref}
\usepackage{natbib}
\usepackage{graphicx,latexsym} 

\begin{document}


\title{Unified trade-off optimization of  quantum harmonic Otto engine and refrigerator} 

\author{Varinder Singh}
\email{varinder@ibs.re.kr}
\affiliation{Center for Theoretical Physics of Complex Systems, Institute for Basic Science (IBS), Daejeon 34126, Korea}
%
\author{Satnam Singh}
\affiliation{ Department of Physical Sciences,   
Indian Institute of Science Education and Research Mohali,
Sector 81, S.A.S. Nagar, Manauli PO 140306, Punjab, India}
\author{Obinna Abah}
\affiliation{Centre for Theoretical Atomic, Molecular and Optical Physics,
School of Mathematics and Physics, Queen’s University Belfast, Belfast BT7 1NN, United Kingdom}
\affiliation{ School of Mathematics, Statistics, and Physics, Newcastle University, Newcastle upon Tyne, NE1 7RU, United Kingdom}%
\author{\"{O}zg\"{u}r E. M\"{u}stecapl{\i}o\u{g}lu}
\affiliation{Department of Physics, Ko\c{c} University, 34450 Sar\i{}yer, Istanbul, Turkey}
\affiliation{TÜBİTAK Research Institute for Fundamental Sciences, 41470 Gebze, Turkey}
%


\begin{abstract}
We investigate quantum Otto engine and refrigeration cycles of a time-dependent harmonic oscillator operating under the conditions of maximum $\Omega$-function, a trade-off objective function which represents a compromise between  energy benefits and losses for a specific job,  for both adiabatic and nonadiabatic (sudden) frequency modulations. We derive analytical expressions for the efficiency and coefficient of performance of the Otto cycle. For the case of adiabatic driving, we point out that in the low-temperature regime, the harmonic Otto engine (refrigerator) can be mapped to Feynman's ratchet and pawl model which is a steady state classical heat engine. For the sudden switch of frequencies, we obtain loop-like behavior of   the efficiency-work curve, which is characteristic of  irreversible heat engines.  Finally, we discuss the behavior of cooling power at maximum $\Omega$-function and indicate the optimal operational point of the refrigerator.
\end{abstract}


\maketitle 

\section{Introduction}
Since the dawn of the industrial revolution, thermal machines have provided the practical impetus to the development of thermodynamics on the experimental and theoretical front. The discovery of Carnot efficiency, which sets a universal upper bound on the efficiency of all heat engines working between two reservoirs, led to the formulation of the second law of thermodynamics by Clausius \cite{DilipBook}. Heat engines and refrigerators are the two well-known  examples of thermal  devices.  Heat engines convert heat energy into useful mechanical work while the refrigerators use external work to lower the temperature of the target system \cite{cen01}. These machines require at least two heat reservoirs at different temperatures, and their performance is limited by the Carnot bound. In the case of heat engines, the Carnot efficiency is given by, $\eta_{\rm C}=1-\beta_2/\beta_1$, where $\beta_1$ ($\beta_2$) is the inverse temperature of the cold (hot) reservoir ($\beta_1>\beta_2$) \cite{cal85,cen01}. 
The corresponding bound on the coefficient of performance (COP) of the refrigerators is given by, $ \zeta_{\rm C}=\beta_2/(\beta_1-\beta_2)$.

However, practical performance of the heat engines/refrigerators are usually lower than the optimal performace due to the associated heat leaks and frictional effects \cite{Insinga:2020,Gordon1991,Gordon1992,Chen2001}. The goal of finite-time thermodynamics is  finding the optimal performance of thermal machines when these limitations are taken into account as well as devising ways to improve on it \cite{Andresen2011,Berry1984,Salamon2001,LevyKosloff}. 
One is usually interested in optimizing the power output of a heat engine and its corresponding efficiency \cite{Curzon1975AJP,Landsberg1989JPA,Lutz2012,Satnam2020,SatnamShishram,Kosloff2017Entropy,SatnamObinna,Myers:2020,Piccione2021PRA,Esposito2009,Esposito2010,MePRR}, whereas, for a refrigerator, the most desirable figure of merit is cooling power \cite{VJ2020,Alonso2014B,LevyKosloff,Apertet2013A}. 
A well-known observation that the thermal engines operating at maximum power also dissipate a large amount of power due to entropy production, which ultimately pollutes the environment \cite{Chen2001,VJ2018,VJ2019}.  Therefore, instead of operating engines (refrigerators) in the maximum power (cooling power) regime, the real irreversible thermal machines should operate near the maximum power point where they yield considerably higher efficiency with a significant reduction in entropy production. Ecological function \cite{ABrown1991}, $\Omega$-function \cite{Hernandez2001}, and efficient power function \cite{Stucki,Yilmaz} are the most commonly studied trade-off objective functions  which pay equal attention to both efficiency and power.
 
Due to its simplicity and amenability to analytical results, a quantum Otto cycle, whose working substance is a time-dependent single harmonic oscillator, has become a standard model to investigate the performance characteristics of  thermal devices \cite{Quan2007,Kieu2004,Rezek2006,Lutz2012,VOzgur2020,Bijay2021,Myers:2020,Hernandez2004,Barish2021,Vahid2021,Assis2019,Assis2020A,Tanmoy2021}. Further, the recent experimental realization of a nanoscale harmonic Otto heat engine provides us with better motivation to study its thermodynamic performance in great detail \cite{Klaers2017}. Although there have been some studies \cite{Rezek2006,Lutz2012,Abah2016EPL} investigating the optimal performance of harmonic Otto heat engines/refrigerators, many aspects remained to be explored, such as performance analysis in the low-temperature regime where both the reservoirs are at low temperatures. Furthermore, an  analytic expression for the COP of the refrigerator is still missing in the sudden limit of operation.

This paper explores the  optimization of $\Omega$-function for the Otto cycle, whose working substance is a quantum harmonic oscillator. In particular, the $\Omega$-function allows a unified trade-off between useful energy delivered and energy lost for heat engines and refrigerators \cite{Hernandez2001,Hernandez2004}, which makes it an ideal figure of merit to study optimal performance of both engines and refrigerators on equal footing.%
We carry out an extensive analysis of the two extreme limiting cases of operation of the Otto cycle: adiabatic limit, which corresponds to quasistatic expansion/compression strokes, and sudden limit of expansion/compression strokes. In both cases, we obtain analytic results for the efficiency (COP) at maximum $\Omega$-function (MOF) of the heat engine (refrigerator).

The paper is organized as follows. In Sec.~II we discuss the model of a harmonic Otto cycle coupled to two thermal reservoirs at different temperatures. Sec.~III presents analytic expressions for the efficiency at maximum $\Omega$-function (EMOF) for both adiabatic and nonadiabatic frequency driving in high and low temperature limits. We also show the loop-like behavior of efficiency-work curve.  In Sec.~IV we repeat the same analysis for the refrigerator cycle. We conclude in Sec.~V.

\section{Quantum Otto cycle}
The quantum Otto cycle consists  two adiabatic and two isochoric thermodynamic processes.  These four steps occur in the following order \cite{Lutz2012,Abah2016EPL}:
(1) Adiabatic compression $A\longrightarrow B$:  Initially, we assume the system is thermalized at 
inverse temperature $\beta_1$. Then, the system is isolated from the environment and the frequency of the oscillator is changed from $\omega_1$ to $\omega_2$ via an external driving protocol. The average energy of the system increases the work being done on the system. The evolution is unitary, and the von Neumann entropy of the system remains constant.
(2) Hot isochore $B\longrightarrow C$: During this stage, the harmonic oscillator is in contact with the hot bath at 
inverse temperature $\beta_2$, and frequency ($\omega_2$) of the oscillator is kept fixed at a fixed value. The system exchanges energy with the hot bath and attains the same temperature of the hot reservoir.  
(3) Isentropic expansion $C \longrightarrow D$: The system is isolated from the surroundings, and the frequency of the harmonic oscillator is unitarily brought back to its initial value $\omega_1$. Work is done by the system in this stage.   
(4) Cold isochore $D\longrightarrow A$: To bring back the working fluid (harmonic oscillator) to its initial state, the system is placed in contact with the cold reservoir at inverse temperature $\beta_1$ ($\beta_1<\beta_2$) at fixed frequency $\omega_1$, and is allowed to relax back to the initial thermal state $A$.

 The average  energies $\langle H \rangle$ of the oscillator at the four stages of the cycle are \cite{Lutz2012,Abah2016EPL} 

\begin{equation}
\langle H\rangle_A =\frac{ \omega_1}{2}\text{coth}\Big(\frac{\beta_1  \omega_1}{2}\Big) ,
\end{equation}
\begin{equation}
\langle H\rangle_B =\frac{ \omega_2}{2} \lambda \text{coth}\Big(\frac{\beta_1  \omega_1}{2}\Big) ,
\end{equation}
\begin{equation}
\langle H\rangle_C =\frac{ \omega_2}{2} \text{coth}\Big(\frac{\beta_2   \omega_2}{2}\Big) ,
\end{equation}
\begin{equation}
\langle H\rangle_D =\frac{ \omega_1}{2}\lambda \text{coth}\Big(\frac{\beta_2  \omega_2}{2}\Big) ,
\end{equation}
where we have set $\hbar=k_B=1$ for simplicity. $\lambda$  is the dimensionless adiabaticity parameter (In the previous works, Refs. \cite{Deffner2008,Lutz2012,Abah2016EPL}, symbol $Q$ is used for adibaticity parameter instead of $\lambda$). The general form of $\lambda$ is given in  Ref. \cite{Deffner2008}. In this paper, we 
will discuss two extreme cases: the adiabatic and sudden switch of frequencies. 
For the  adiabatic process, $\lambda=1$ and for the sudden switch of frequencies,  
$\lambda=(\omega_1^2+\omega_2^2)/2 \omega_1 \omega_2$ \cite{{Lutz2012,Abah2016EPL}}.
The expression for  mean heat exchanged during the hot and cold isochores can be evaluated, respectively, as follows
 \begin{eqnarray}
   Q_2   &=& \langle H \rangle_C-\langle H \rangle_B \nonumber
  \\
  &=& \frac{ \omega_2}{2}\Big[\text{coth}\Big(\frac{\beta_2 \omega_2}{2}\Big)-\lambda\text{coth}\Big(\frac{\beta_1  \omega_1}{2}\Big) \Big]  \label{heat2}
 \end{eqnarray}
  \begin{eqnarray}
  Q_4   &=& \langle H \rangle_A-\langle H \rangle_D \nonumber
  \\
  &=& \frac{  \omega_1}{2}\Big[\text{coth}\Big(\frac{\beta_1 \omega_1}{2}\Big)-\lambda\text{coth}\Big(\frac{\beta_2\omega_2}{2}\Big) \Big]. \label{heat4}   
 \end{eqnarray}
We are employing a sign convention in which all the incoming fluxes (heat and work) are taken to be positive.
  \begin{figure}
 \centering
 \includegraphics[scale=0.65,keepaspectratio=true]{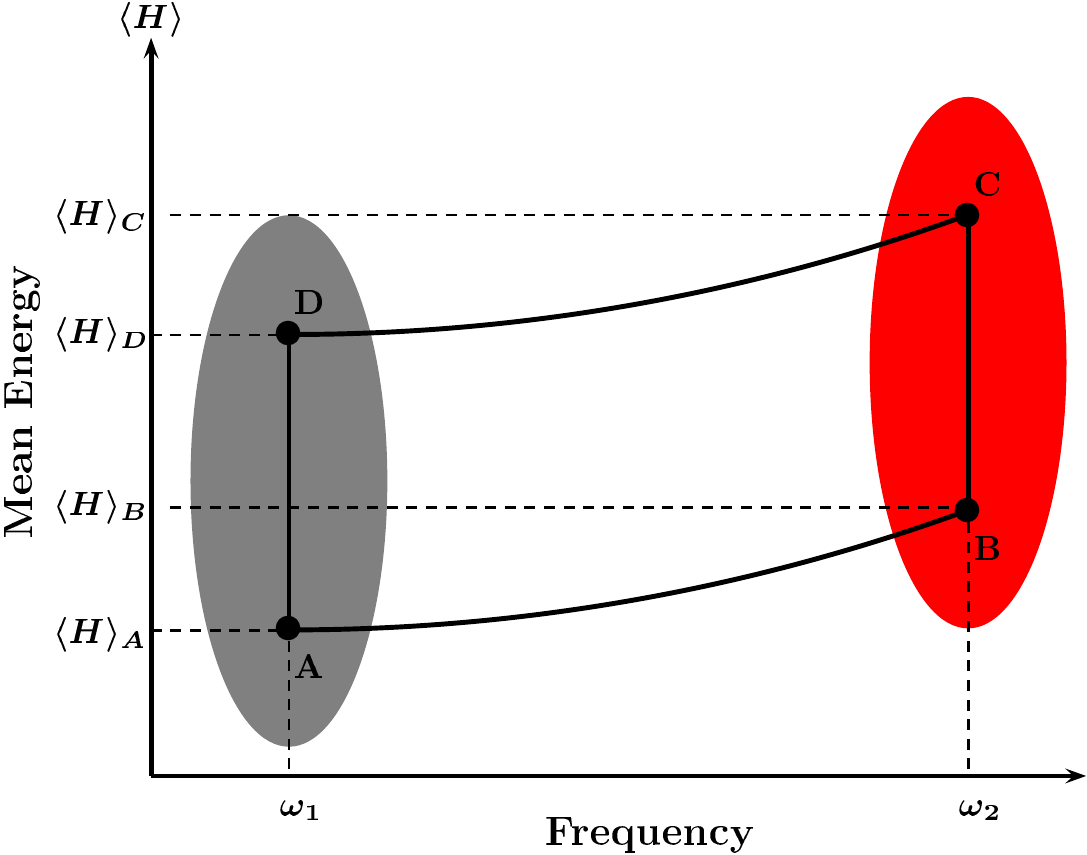}
 \caption{Pictorial depiction of  Otto cycle. The thermodynamic cycle consists of four stages: two adiabatic (A$\rightarrow$ B and C$\rightarrow$ D) and two isochoric (B$\rightarrow$ C and D $\rightarrow$ A) steps.} \label{fig:cycle}
\end{figure}
\section{Quantum Otto heat engine}
Since  the working fluid returns to its initial state after one complete cycle, the extracted work in one 
complete cycle is given by $W=Q_2+Q_4>0$. Accordingly, the efficiency of the engine is given by
\begin{equation}
\eta = \frac{W}{Q_2} 
=
1-\frac{\omega_1}{\omega_2}\frac{\text{coth}(\beta_1   \omega_1/2)-\lambda\text{coth}(\beta_2\omega_2 /2)}{\lambda \text{coth}(\beta_1  \omega_1/2)-\text{coth}(\beta_2  \omega_2 /2)}. \label{eq:efficiency-general}
\end{equation}

The optimal performance of the harmonic Otto engine at maximum work/power has been studied already \cite{Lutz2012}. In this work, we optimize the Omega function, which represents a compromise between the useful work and the loss of work in the system \cite{Hernandez2001}. It is defined as follows \cite{Hernandez2001}
\begin{equation}
\Omega = 2 W - \eta_{\rm max} Q_2 \label{Omega},
\end{equation}
where $\eta_{\rm max}\leq \eta_{\rm C}$ is the maximum possible efficiency achievable to the engine under consideration. The $\Omega$-function is equivalent to an another trade-off function known as ecological function when $\eta_{\rm max}=\eta_{\rm C}$ \cite{ABrown1991,Hernandez2001}.  

 For the harmonic Otto cycle, $\eta_{\rm max}$ depends on the speed of the adiabatic protocol which is expressed in terms $\lambda$ \cite{VOzgur2020}.
We will show in a moment that in the adiabatic case, $\eta_{\rm max}=\eta_{\rm C}$. However, in the case of nonadiabatic work strokes, the maximum efficiency of the engine under consideration is  always less than the Carnot efficiency due to internal friction. Particularly, for the sudden-switch case, the maximum efficiency of the engine is given by Eq. (\ref{etamax}) \cite{VOzgur2020}. %
In the following, first we will discuss the adiabatic case and then move on to discuss nonadiabatic case.
\subsection{Adiabatic case}
Quantum adiabatic  processes are much slower than the typical time scales of the system. In this case,
$\lambda=1$, and from the positive work condition ($W=Q_2+Q_4>0$), we find $\eta_{\rm max}=\eta_{\rm C}$.  Using Eqs.(\ref{eq:efficiency-general}) and (\ref{Omega}), the expressions for efficiency and $\Omega$-function take the  forms
\begin{equation}
 \eta=1-\frac{\omega_1}{\omega_2} , \qquad \Omega = 2W -\eta_{\rm C} Q_2. \label{effOmega}
 \end{equation}
\subsubsection{High-temperature regime}
In order to obtain analytic expression in closed form for the  EMOF, we
will work in the high temperature regime. In the high-temperature regime, we set
$\coth(\beta_i\omega_i/2)\approx 2/(\beta_i\omega_i)$ ($i=1, 2$). 
Using Eqs. (\ref{heat2}) and (\ref{heat4})   in Eq. (\ref{effOmega}), the expression for $\Omega$-function is written as
\begin{equation}
\Omega = \frac{(z-\tau)(1+\tau-2z)}{\beta_2 z}, \label{eco3}
\end{equation}
where $z\equiv \omega_1/\omega_2$ is the compression ratio of the Otto cycle, and $\tau=\beta_2/\beta_1$. Optimization of Eq. (\ref{eco3}) 
with respect to compression ratio $z$ yields $z^* = \sqrt{\frac{\tau(1+\tau)}{2}}$. Hence, EMOF, in terms of Carnot 
efficiency, is given by

\begin{equation}
\eta^{\Omega}_{\rm high} = 1- \sqrt{\frac{(1-\eta_{\rm C})(2-\eta_{\rm C})}{2}}, \label{AB}
\end{equation}
which concurs with the efficiency of the endoreversible and symmetric low-dissipation models of heat engines \cite{ABrown1991,deTomas2013}. The results are not surprising as in the high-temperature regime (classical regime), the engines are expected to behave like classical heat engines \cite{Geva1992,VJ2019,VJ2020}. Eq. (\ref{AB}) was first obtained by Angulo-Brown for the optimization of endoreversible heat engines \cite{ABrown1991}.The corresponding efficiency for the optimization of the harmonic Otto engine operating under the conditions of maximum work output is given by Curzon-Ahlborn formula, $\eta^{W}_{\rm high}=\eta_{\rm CA}=1-\sqrt{1-\eta_{\rm C}}$ \cite{Lutz2012,Rezek2006}. See Eqs. (\ref{taylor2}) and (\ref{T1}) for the comparison of   $\eta^{W}_{\rm high}$ and $\eta^{\Omega}_{\rm high}$. It is clear that $\eta^{\Omega}_{\rm high}$ is always greater than  $\eta^{W}_{\rm high}$, which is expected outcome \cite{ABrown1991,Hernandez2001}.
\subsubsection{Low-temperature regime}
Here, we will discuss the performance of the harmonic Otto engine in the low-temperature regime which has not 
been explored in earlier publications. In Refs. \cite{Lutz2012,Abah2016EPL,Abah2014EPL}, the optimization has been carried out in the regime 
defined by the constraints $\beta_1\omega_1\gg 1$ and $\beta_2\omega_2\ll 1$, i.e., the hot reservoir being very hot 
and the cold reservoir being very cold. In the following, we will discuss adiabatic case only as it is not possible to 
obtain analytic results for the non-adiabatic case. We assume that $\beta_i\omega_i\gg 1$, and set $\coth(\beta_{i}\omega_{i}/2)\approx 1+2e^{-\beta_{i}\omega_{i}}$. Using Eqs. (\ref{heat2})
and  (\ref{heat4}) in the expression, $W=Q_2+Q_4$, the extracted work, in the low-temperature limit, can be 
expressed as follows:
\begin{equation}
W = (\omega_2-\omega_1)\left( e^{-\beta_2\omega_2} -e^{-\beta_1\omega_1}   \right). \label{FR}
\end{equation}
Apart from a multiplicative constant, the above expression for extracted work is similar to the expression for 
the power output of the Feynman's ratchet and pawl model, where control parameters are internal energy states 
$\epsilon_1$ and $\epsilon_2$ instead of $\omega_1$ and $\omega_2$ \cite{Tu2008,Feynman}. Thus in the low-temperature limit, the harmonic 
Otto engine can be mapped to Feynman's model, which is a steady-state classical heat engine based on the principle 
of  Brownian fluctuations \cite{Feynman,Tu2008}. Interestingly, it is not the only case in which a quantum heat engine can be mapped to Feynman's ratchet and pawl engine. Recently, a three-level laser quantum heat engine operating
in the low-temperature regime was also mapped to Feynman's model \cite{VJ2019}.  

In order to obtain the analytic expression for the efficiency which is independent of 
the parameters of the system and depends on the ratio of bath temperatures only, the optimization should be carried 
out with respect to two variables ($\omega_1$ and $\omega_2$) simultaneously.  Treating $\omega_1$ and $\omega_2$ 
as the independent variables,  optimization of  Eq. (\ref{FR}) with respect to  $\omega_1$ and $\omega_2$ yields the
following optimal solution \cite{Tu2008}
\begin{eqnarray}
\omega_1^* &=& \frac{(1-\eta_{\rm C})[\eta_{\rm C}-\ln(1-\eta_{\rm C})]}{\eta_{\rm C} \beta_2},\label{FRsolution1}
\\
\omega_2^* &=& \frac{\eta_{\rm C}-(1-\eta_{\rm C})\ln(1-\eta_{\rm C})}{\eta_{\rm C} \beta_2}. \label{FRsolution2}
\end{eqnarray}
 Using 
Eqs. (\ref{FRsolution1}) and (\ref{FRsolution2})   in Eqs. (\ref{effOmega}) and (\ref{FR}), we obtain the expressions for the EMOF and optimal work, respectively.
\begin{equation}
\eta^{W}_{\rm low} = \frac{\eta_{\rm C}^2}{\eta_{\rm C}-(1-\eta_{\rm C})\ln(1-\eta_{\rm C})}, \label{effFR}
\end{equation}
\begin{equation}
W^*_{\rm low} = \frac{\eta_{\rm C}^2(1-\eta_{\rm C})^{(1-\eta_{\rm C})/\eta_{\rm C}}}{\beta_2 e} .
\end{equation}

Using these analytic expressions, we discuss the universal nature of efficiency at maximum work (EMW). For near equilibrium conditions, 
expanding ($\beta_1\approx\beta_2$) Eq. (\ref{effFR}) in Taylor series, we have
\begin{equation}
\eta^{W}_{\rm low} = \frac{\eta_{\rm C}}{2} + \frac{\eta_{\rm C}^2}{8} + \frac{7\eta_{\rm C}^3}{96} + \mathcal{O}(\eta_{\rm C}^4). \label{taylor1}
\end{equation}
For comparison, we also present the Taylor series expansion of $\eta^W_{\rm high}$,
\begin{equation}
\eta^{W}_{\rm high} = \frac{\eta_{\rm C}}{2} + \frac{\eta_{\rm C}^2}{8} + \frac{6\eta_{\rm C}^3}{96} + \mathcal{O}(\eta_{\rm C}^4). \label{taylor2}
\end{equation}
Notice that $\eta^{W}_{\rm low}>\eta^{W}_{\rm high}$. The first two terms in both Eqs. (\ref{taylor1}) and (\ref{taylor2}) are $\eta_{\rm C}/2$ and $\eta_{\rm C}^2/8$, and third term
is model dependent. For heat engines obeying tight-coupling condition (no heat leaks), universality of  first
term $\eta_{\rm C}/2$ was proven by Van den Broeck using the formalism of linear irreversible thermodynamics \cite{Broeck2005}. 
Further, the universality of second term can be proved by invoking the symmetry of Onsager coefficients on the 
nonlinear level \cite{Esposito2009}.  

Similarly, for the optimization of $\Omega$ function, the optimal solution is given by \cite{VJ2019}
\begin{equation}
\omega_1^{\Omega} = \frac{[\eta_{\rm C}+(2-\eta_{\rm C})k]}{\beta_1 \eta_{\rm C}},
\quad
\omega_2^{\Omega} = \frac{[\eta_{\rm C}+2k(1-\eta_{\rm C})]}{\beta_1 \eta_{\rm C}(1-\eta_{\rm C})}, \label{optimalsol}
\end{equation}
where $k=\ln[(2-\eta_{\rm C})/2(1-\eta_{\rm C})]$. We obtain  efficiency, $\eta=1-\omega_c/\omega_h$, at MOF as follows
\begin{equation}
\eta^{\Omega}_{\rm low}  = \frac{\eta_{\rm C} + (1-\eta_{\rm C})k} {\eta_{\rm C} + 2(1-\eta_{\rm C})k}\, \eta_{\rm C}, \label{lowFR}
\end{equation}
\begin{equation}
\Omega^{*}_{\rm low} = \frac{\eta_{\rm C}^2[2(1-\eta_{\rm C})]^{2(1-\eta_{\rm C})/\eta_{\rm C}}}{\beta_2 e (2-\eta_{\rm C})^{(2-\eta_{\rm C})/\eta_{\rm C}} }.
\end{equation}
Similar to the case of work optimization, $\eta^{\Omega}_{\rm low}$ is independent of the parameters of the system (harmonic oscillator here)
and depends on ratio of reservoir temperatures ($\beta_2/\beta_1$) only. Now, we turn to the universal nature of 
efficiency. The universal nature of efficiency is not unique feature of the optimization of work/power output of 
the engine, the EMOF also shows universal behaviour \cite{Zhang2016}.  Taking near-equilibrium series expansions of Eqs. (\ref{AB}) 
and (\ref{lowFR}), we have
\begin{eqnarray}
\eta^{\Omega}_{\rm high} &=& \frac{3\eta_{\rm C}}{4} + \frac{\eta_{\rm C}^2}{32} + 
\frac{18\eta_{\rm C}^3}{768} + {\cal O}(\eta_{\rm C}^4), \label{T1}
\\
\eta^{\Omega}_{\rm low} &=& \frac{3\eta_{\rm C}}{4} + \frac{\eta_{\rm C}^2}{32} + 
\frac{19\eta_{\rm C}^3}{768} + {\cal O}(\eta_{\rm C}^4). \label{T2}
\end{eqnarray}
Again, it is self evident from Eqs. (\ref{T1}) and (\ref{T2}) that first two terms of $\eta^{\Omega}_{\rm high}$ and 
$\eta^{\Omega}_{\rm low}$ are same, and the model dependent differences appear in the third term only. 
The universality of the EMOF was first formally proven by Zhang and 
coauthors \cite{Zhang2016} by using the framework of stochastic thermodynamics. The first two terms 
$3\eta_{\rm C}/4$ and $\eta_{\rm C}^2/32$ were also obtained for  endoreversible \cite{ABrown1991,Sanchez2010}, low-dissipation \cite{deTomas2013}, minimally nonlinear irreversible \cite{LongLiu2014} and some other models of classical and quantum heat engines \cite{LongLiu2015_2,Sanchez2010,VJ2019}.

\subsection{Sudden switch of frequencies}
Next, we discuss the case in which frequency of the oscillator is changed suddenly from
one value to the other. In this case, $\lambda=(\omega_1^2+\omega_2^2)/2 \omega_1 \omega_2$ \cite{Deffner2008}. The efficiency is no longer given by Eq. (\ref{effOmega}), and expression for the efficiency, in the high-temperature limit, reads as
\begin{equation}
\eta_{\rm SS} = \frac{\left(z^2-1\right) \left(z^2-\tau \right)}{\tau +(\tau -2) z^2}. \label{effss}
\end{equation}
Similarly, the expression for the work output is given by,
\begin{equation}
W_{\rm SS} =  \frac{(1-z^2)(z^2-\tau)} {2z^2 \beta_2}. \label{Wss}
\end{equation}
From the positive work condition, $W_{\rm SS}>0$, we have 
\begin{equation}
z^2>\tau \Rightarrow z >\sqrt{\tau}. \label{pwc}
\end{equation}
The above condition is more restrictive than the positive work condition for the adiabatic case which implies that $z>\tau$.  Hence, for the given temperatures of the cold and hot reservoirs, it is more difficult to extract work for the sudden switch case as compared to the adiabatic one.

Here, we are interested in the optimization of $\Omega$ function. In order to find the expression for $\Omega$ function, first we have to specify $\eta_{\rm max}$. Recently, the form of $\eta_{\rm max}$
is evaluated in Ref. \cite{VOzgur2020}, and,  is given by
\begin{equation}
\eta^{\rm SS}_{\rm max} = \frac{\big[ 3-\eta_{\rm C}-2\sqrt{2(1-\eta_{\rm C})}  \big]\eta_{\rm C}}{(1+\eta_{\rm C})^2} \leq \frac{1}{2}. \label{etamax}
\end{equation}
Substituting Eq. (\ref{etamax}) in Eq. (\ref{effOmega}), we obtain required expression for the $\Omega$-function for the nonadiabatic
(sudden-switch) case as follows:
\begin{widetext}
\begin{equation}
\Omega_{SS} =  \frac{\left(1-z^2\right) \left(2 \left(z^2-\tau \right)+\frac{(\tau -1) \left(\tau -2 \sqrt{2} \sqrt{\tau }+2\right) \left(\tau +(\tau -2) z^2\right)}{(\tau -2)^2 \left(z^2-1\right)}\right)}{2 z^2}.
\end{equation}
\end{widetext}
Then by optimizing the $\Omega$-function, EMOF is found to be: 
\begin{equation}
\eta^{\Omega}_{SS}= \frac{(2+2\eta_{\rm C}-A)(2-2\eta_{\rm C}^2-A)}{2(1+\eta_{\rm C})^2(2-2\eta_{\rm C}-A)}, \label{EMOF}
\end{equation}
where $A=\sqrt{2(1-\eta_{\rm C})\big(2+\eta_{\rm C}+2\eta_{\rm C}\sqrt{2(1-\eta_{\rm C})}+3\eta_{\rm C}^2\big)}$.
This efficiency can be considered as the counterpart of the EMW, 
$\eta^{W}_{\rm SS}=(1-\sqrt{1-\eta_{\rm C}})/(2+\sqrt{1-\eta_{\rm C}})$, which was obtained for the optimization of 
the work output of the harmonic Otto engine undergoing sudden compression and expansion strokes during the adiabatic branches  \cite{Rezek2006}. 
\begin{figure}   
	\begin{center}
		\includegraphics[width=8.6cm]{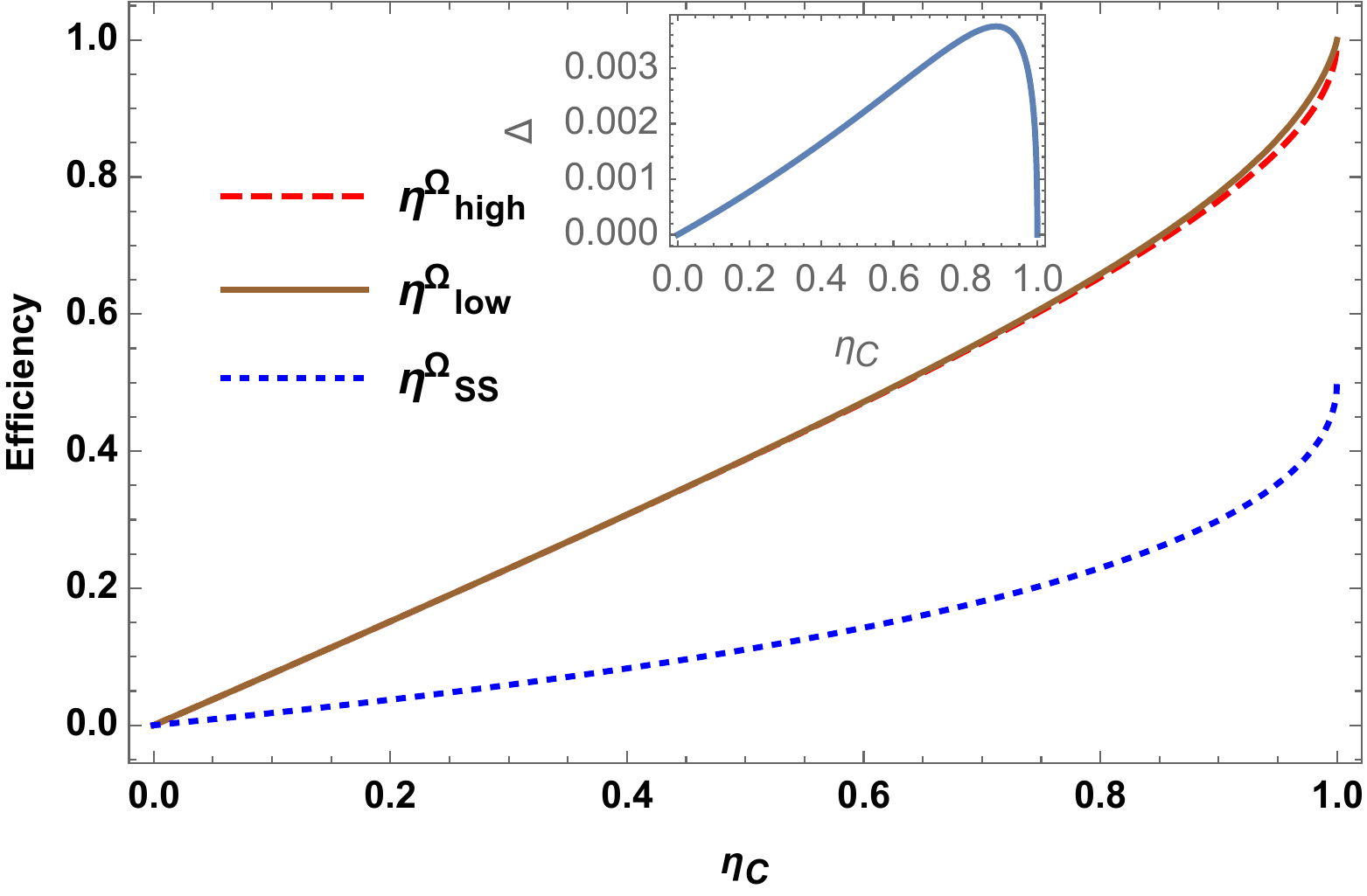}
	\end{center}
	\caption{EMOF for the adiabatic (upper set of curves) and sudden-switch (lowest lying curve) cases versus Carnot efficiency. Solid brown, dashed red and dotted blue curves represent  Eqs. (\ref{AB}), (\ref{lowFR}) and   (\ref{EMOF}),  respectively.
In the inset, we have plotted the difference ($\Delta$) between $\eta^\Omega_{SS}$ and $\eta^W_{SS}$. All the notations used here are defined in the main text.} \label{comparison}
\end{figure}

In order to compare the performance of the engine for the adiabatic driving and sudden switch case, we plot Eqs. (\ref{AB}) and  (\ref{lowFR}) along with the expressions for Eq. (\ref{EMOF}) in Fig. \ref{comparison}.
In the inset of Fig. 2, we have plotted the difference between $\eta^W_{SS}$ and   $\eta^\Omega_{SS}$. Although in the sudden switch case, EMOF is larger than EMW, the difference is not substantial. 
Further, the    EMOF is very low in the sudden switch regime as compared to the adiabatic driving.
We attribute this to the highly frictional nature of the sudden switch regime as explained below. In the 
sudden switch case, the sudden change of the frequency of the harmonic oscillator induces nonadiabatic 
transitions between its energy levels and leaves the system in a highly nonequilibrium state. In terms of the 
energy eigenstates of the instantaneous Hamiltonian, the off-diagonal terms of the density matrix, known as coherences, 
are non-zero.  Generating coherences give rise to extra energetic cost when compared to adiabatic driving, 
and an additional parasitic internal energy is stored in the working medium.
This extra cost gets  dissipated to the heat reservoirs
during the proceeding isochoric stages  of the cycle, and is termed as quantum friction \cite{Rezek2010,Plastina2014,Rezek2017,Feldmann2000,Feldmann2006,Ozgur2017}. Inner friction is detrimental for the performance of the engine under consideration.

\begin{figure}
 \begin{center}
 \includegraphics[width=8.6cm]{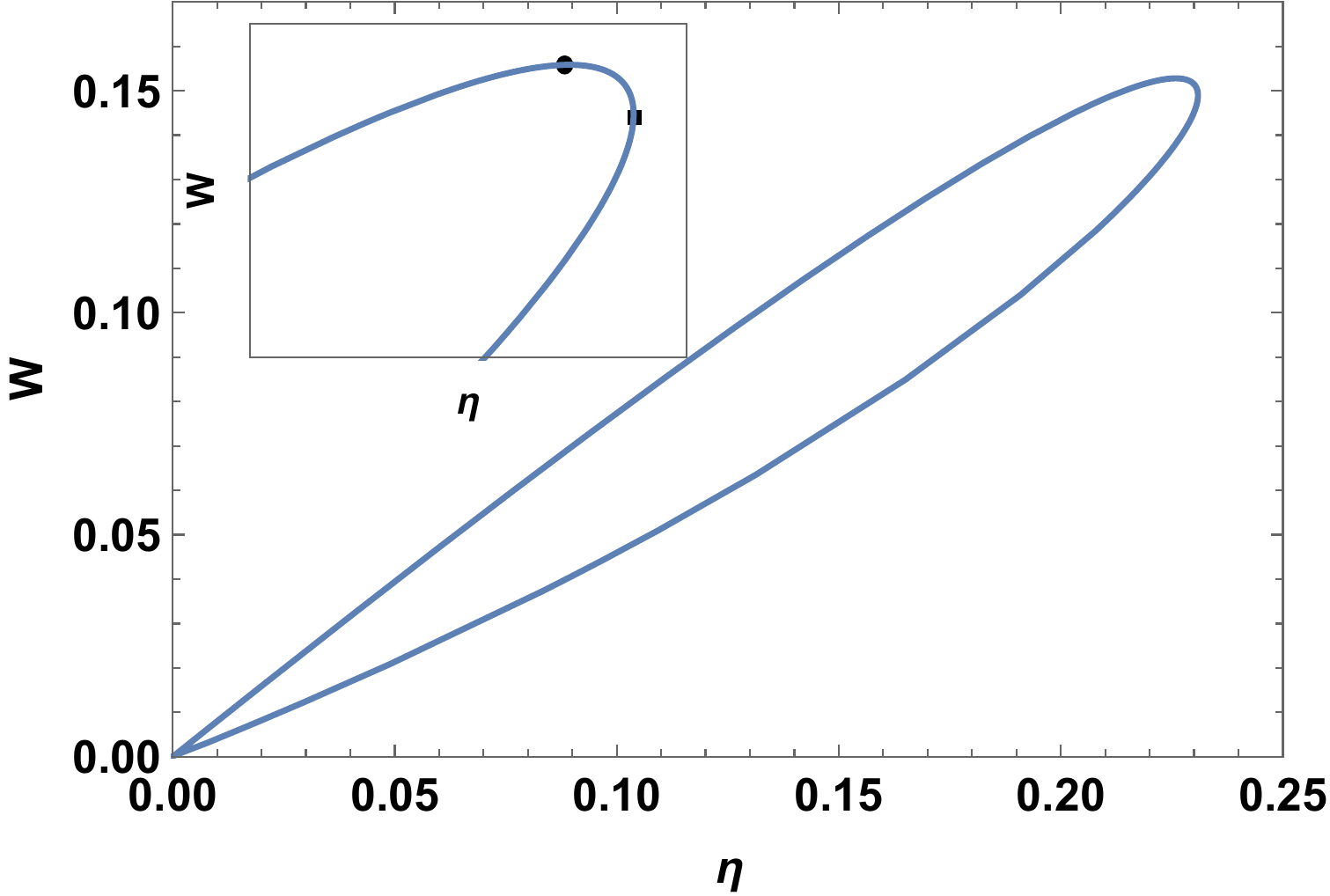}
 \end{center}
 \caption{Loop shaped work versus efficiency curve, characteristic of real irreversible heat engines. In the inset, we have plotted the amplified version of efficiency-work curve to distinguish the maximum work and maximum efficiency points. Maximum efficiency (black dot) and maximum work (square dot) points lie very close to each other.} \label{loop}
\end{figure}

Now, we plot typical efficiency-work   curves in Fig. 3. Using Eqs. (\ref{effss}) and (\ref{Wss}), the parametric plot between work and efficiency is obtained (see Fig. \ref{loop}).
The  efficiency-work ($\eta, W$) curve shows the
loop-like behavior,  characteristic of realistic irreversible heat engines \cite{Gordon1991,Gordon1992,GordonKosloff,Benenti2017}.  As shown in Fig. \ref{loop}, the maximum efficiency and maximum work output points lie extremely close to each other. The optimal operating regime of the nonadiabatic engine under consideration is situated on the part of ($\eta, W$) curve, which has a negative slope, i. e., the portion of the ($\eta, W$) curve lying in between maximum work and maximum efficiency point. The optimization of the $\Omega$ function lies in this regime. It is worth mentioning that the loop shape of the work efficiency curve arises due to the presence of inner friction in the operation of the engine which is a purely quantum mechanical effect, as mentioned earlier. The loop-like behavior can also be seen in power-efficiency curve of classical endoreversible heat engines in the presence of heat-leaks in the system \cite{Chen1994,Chen2001}. 

Further, the loop-shape ($\eta, W$) curves are not exclusive to the sudden adiabatic strokes (the case considered here). They can be obtained for any adiabatic stroke happening in finite time, thus giving rise to nonadiabatic transitions between the energy levels of the harmonic oscillator, which are responsible for the appearance of inner friction. As time spent on adiabatic branches increases, the maximum work and maximum efficiency points on the ($\eta, W$)  curve will move further apart. Finally, for the quasistatic process, the  ($\eta, W$)  curve will become open parabolic in shape, just like the efficiency-power curve of endoreversible heat engines without heat leaks \cite{Gordon1991,Gordon1992,Chen2001}.

\section{Quantum Otto refrigerator}
In this section, we investigate the performance of the harmonic Otto cycle working as a refrigerator in the adiabatic as well as non-adiabatic regime. For the refrigerator, $Q_4>0$, $Q_2<0$, and the work invested to transport heat from the cold reservoir to the hot reservoir is positive, $W_{\rm in}=- (Q_2+ Q_4)= W_1+ W_3>0$. The coefficient of performance of the refrigerator is defined 
by
\begin{equation}
\zeta = \frac{Q_4}{W_{\rm in}}=-\frac{Q_4}{Q_2+ Q_4}. 
\label{cop}
\end{equation}
Using Eqs. (\ref{heat2}) and (\ref{heat4}) in Eq. (\ref{cop}), the COP takes the following form \cite{Abah2016EPL}:
\begin{equation}
\zeta = \frac{\omega_1[\coth(\beta_1\omega_1/2) - \lambda\coth(\beta_1\omega_1/2)]}{(\lambda\omega_2-\omega_1)\coth(\beta_1\omega_1/2)-(\omega_2-\lambda\omega_1)\coth(\beta_2\omega_2/2)}. \label{cop22}
\end{equation}

The $\Omega$ function for the refrigerator is given by \cite{Hernandez2001}
\begin{equation}
\Omega = 2 Q_4  - \zeta_{\rm max}  W_{\rm in}, \label{omegaref1}
\end{equation}
where $\zeta_{\rm max}\leq \zeta_{\rm C}$ is the maximum COP with which our refrigerator can operate. It is different for adiabatic and nonadiabatic cases. As for the case of heat engine, when $\zeta_{\rm max}= \zeta_{\rm C}$, the $\Omega$ function is equivalent to ecological function \cite{Yan_1996} defined for refrigerators. As before, first we will discuss adiabatic case. 
\subsection{Adiabatic driving}
Substituting $\lambda=1$ in Eq. (\ref{cop22}), the form of is found to be
\begin{equation}
\zeta_{\rm ad} = \frac{\omega_1}{\omega_2-\omega_1}=\frac{z}{1-z}. \label{cop2}
\end{equation}
Similarly, substituting $\lambda=1$ in Eq. (5), the positive cooling condition, $Q_4>0$, implies that 
$\beta_1\omega_1<\beta_2\omega_2$, which in turn implies that $\zeta< \zeta_{\rm C}$. Hence, for the adiabatic driving, 
$\zeta_{\rm max}= \zeta_{\rm C}$. Therefore, from Eq. (\ref{omegaref1}), we have
\begin{equation}
\Omega = 2  Q_4   - \zeta_{\rm C}   W_{\rm in} . \label{Omegaref}
\end{equation}
\subsubsection{High-temperature regime}
Again, to evaluate analytic expressions for the cop, we choose to work in the high-temperature regime. In this regime, 
using Eqs. (\ref{heat2}), (\ref{heat4}) along with $W_{\rm in}=-(Q_2+Q_4)$, $\Omega$ function can be written in terms of $\tau$ and $z$ and is given by:
\begin{equation}
\Omega =  \frac{(z-\tau)[\tau-z(2-\tau)]}{\beta_2 z(1-\tau)}.
\end{equation}
Optimization of $\Omega$ with respect to $z$ yields the following optimal solution,
\begin{equation}
z^* = \frac{\tau}{\sqrt{2-\tau}}. \label{zadoptimal}
\end{equation}
Substiting Eq. (\ref{zadoptimal}) in Eq. (\ref{cop2}), we have
\begin{equation}
\zeta^{\Omega}_{\rm high}= \frac{\tau }{\sqrt{2-\tau }-\tau }=\frac{ \zeta_{\rm C}}{\sqrt{(1+ \zeta_{\rm C})(2+ \zeta_{\rm C})}- \zeta_{\rm C}}. \label{optimalcophigh}
\end{equation}
The above expression for the COP concurs with those of the endoreversible \cite{Yan_1996} and symmetric low-dissipation models of heat engines \cite{deTomas2013}.  The corresponding  COP at maximum $\chi$-criterion, which is the product of the COP and CP of a refrigerator, is given by the formula, $\zeta^{\chi}_{\rm high}=\sqrt{1+ \zeta_{\rm C}}-1$ \cite{Abah2016EPL}.  It is worth to mention that $\zeta^{\chi}_{\rm high}$ is always lower than $\zeta^{\Omega}_{\rm high}$.%

\subsubsection{Low-temperature regime}
In parallel to the heat engine description, here we investigate the performance of the refrigerator in the 
low-temperature regime. In the low-temperature regime, the expressions for $Q_4$ and 
$\Omega$ take the forms:
\begin{equation}
Q_4 = \omega_1  \left( e^{-\beta_1\omega_1} - e^{-\beta_2\omega_2} \right), \label{coolLT}
\end{equation}
\begin{equation}
\Omega = [(2+ \zeta_{\rm C})\omega_1- \zeta_{\rm C}\omega_2]\left( e^{-\beta_1\omega_1} - e^{-\beta_2\omega_2} \right).
\label{omegacold}
\end{equation}
Performing the two-parameter optimization of Eq. (\ref{omegacold}) with respect to control parameters $\omega_1$
and $\omega_2$, the optimal solution is obtained as \cite{Kiran2021}
\begin{equation}
\omega_1^*= \frac{1-(1+ \zeta_{\rm C})k}{\beta_1},\quad \omega_2^*= \frac{ 1-(2+ \zeta_{\rm C})k}{\beta_2}, \label{coolLTsol}
\end{equation}
where $k=\ln[(1+ \zeta_{\rm C})/(2+ \zeta_{\rm C})]$. Substituting above expressions for $\omega_1$ and $\omega_2$ in Eq. (\ref{cop2}), 
the final expression for the optimal COP is found to be:
\begin{equation}
\zeta^{\Omega}_{\rm low} = \frac{1-(1+ \zeta_{\rm C})k}{1-2(1+ \zeta_{\rm C})k} \zeta_{\rm C}. \label{optimalcoplow}
\end{equation}
Similar to the case of heat engine, COP of the refrigerator does not depend on the system parameters and 
depends on the ratio of the reservoir temperatures ($\tau$) only. As we have shown that the adiabatic harmonic Otto cycle operating in the low temperature regime can be mapped to Feynman's model, the above expression also holds for the optimization  of Feynman's ratchet and pawl model \cite{VarinderJohal} and a three-level laser quantum refrigerator \cite{Kiran2021}. For comparison, Eq. (\ref{optimalcophigh}) (solid yellow curve) and Eq. (\ref{optimalcoplow}) (dotted red curve) are plotted in Fig. \ref{copsgraph} and we note that they are practically indistinguishable for the entire range of the graph. This can be understood by looking at the Taylor series behavior of $\zeta^{\Omega}_{\rm high}$ and $\zeta^{\Omega}_{\rm low}$ near equilibrium:
\begin{eqnarray}
\zeta^{\Omega}_{\rm high}  &=& \frac{2 \zeta_{\rm C}}{3}+\frac{1}{18} -\frac{17}{216 \zeta_{\rm C}}+O\left(\frac{1}{ \zeta_{\rm C}^3}\right),
\\
\zeta^{\Omega}_{\rm low}  &=& \frac{2 \zeta_{\rm C}}{3}+\frac{1}{18} -\frac{16}{216 \zeta_{\rm C}}+O\left(\frac{1}{ \zeta_{\rm C}^3}\right).
\end{eqnarray}
The first two terms of on the right hand side of the above equations are the third term is negligible for any value of
$ \zeta_{\rm C}>1$, thus explaining the overlap of curves representing $\zeta^{\Omega}_{\rm high}$ and $\zeta^{\Omega}_{\rm low}$ in Fig. \ref{copsgraph}.

\subsection{Sudden switch of frequencies}
Next, we discuss the case in which frequency of the oscillator is changed suddenly from
one value to the other. In this case, $\lambda=(\omega_1^2+\omega_2^2)/2 \omega_1 \omega_2$. The optimal performance 
of the harmonic Otto refrigerator operating in the sudden-switch regime has not been fully explored earlier. In Ref. \cite{Abah2016EPL}, 
the performance of the refrigerator was studied for weak nonadiabatic driving (for adiabaticity parameter $\lambda$ 
close to 1), and the analytic expression for corresponding minimal driving time was obtained. In the sudden switch regime, the cooling power does not  exhibit a generic maximum with respect to control parameter $z$, i.e., the maximum of the cooling power is obtained for $z=0$, which is clearly not a useful result. Hence, to study the optimal operation of the refrigerator working under the conditions of MOF is a sensible option. Here, we obtain the analytic expression for the COP at optimal $\Omega$-function.  

For sudden-switch driving protocol, the expression for the cooling power and input work are evaluated to be, %
\begin{equation}
Q_4=\frac{1}{\beta_2}\left[\tau -\frac{1}{2} \left(z^2+1\right)\right], \quad W_{\rm total}= \frac{\left(z^2-1\right) \left(z^2-\tau \right)}{2 \beta_2 z^2}.
\label{coolingSS}
\end{equation}
Further, the COP, $ \zeta_{\rm C}=Q_4/W_{\rm in}$, takes the form,
\begin{equation}
\zeta_{ss} = \frac{z^2 \left(2 \tau -z^2-1\right)}{\left(z^2-1\right) \left(z^2-\tau \right)} \label{copss}.
\end{equation}
In order to proceed further, we have to specify form of maximum COP.  Recently, in Ref. \cite{VOzgur2020}, using the positive cooling condition, it was shown that in the sudden-switch regime, the maximum COP  of the harmonic Otto refrigerator is no longer given by Carnot COP $ \zeta_{\rm C}$. The desired form of maximum COP, which is much tighter than the Carnot bound,  is found to be \cite{VOzgur2020},
\begin{equation}
\zeta_{\rm max} = 1 + 3 \zeta_{\rm C} -2\sqrt{2 \zeta_{\rm C}(1+ \zeta_{\rm C})}. \label{copmax}
\end{equation} 
Another interesting constraint imposed by the condition $Q_c>0$ on the temperatures of the reservoirs is:
 \begin{equation}
 \tau > \frac{1}{2} \qquad \text{or} \qquad  \zeta_{\rm C}>1. \label{constraint}
 \end{equation}
The above condition has interesting implications on the performance of the Otto refrigerator operating in the sudden switch regime. Eq. (\ref{constraint}) simply implies that the thermal machine under consideration cannot work as a refrigerator unless the temperature of the cold reservoir is greater than $T_2/2$.
\begin{figure}
 \begin{center}
 \includegraphics[width=8.6cm]{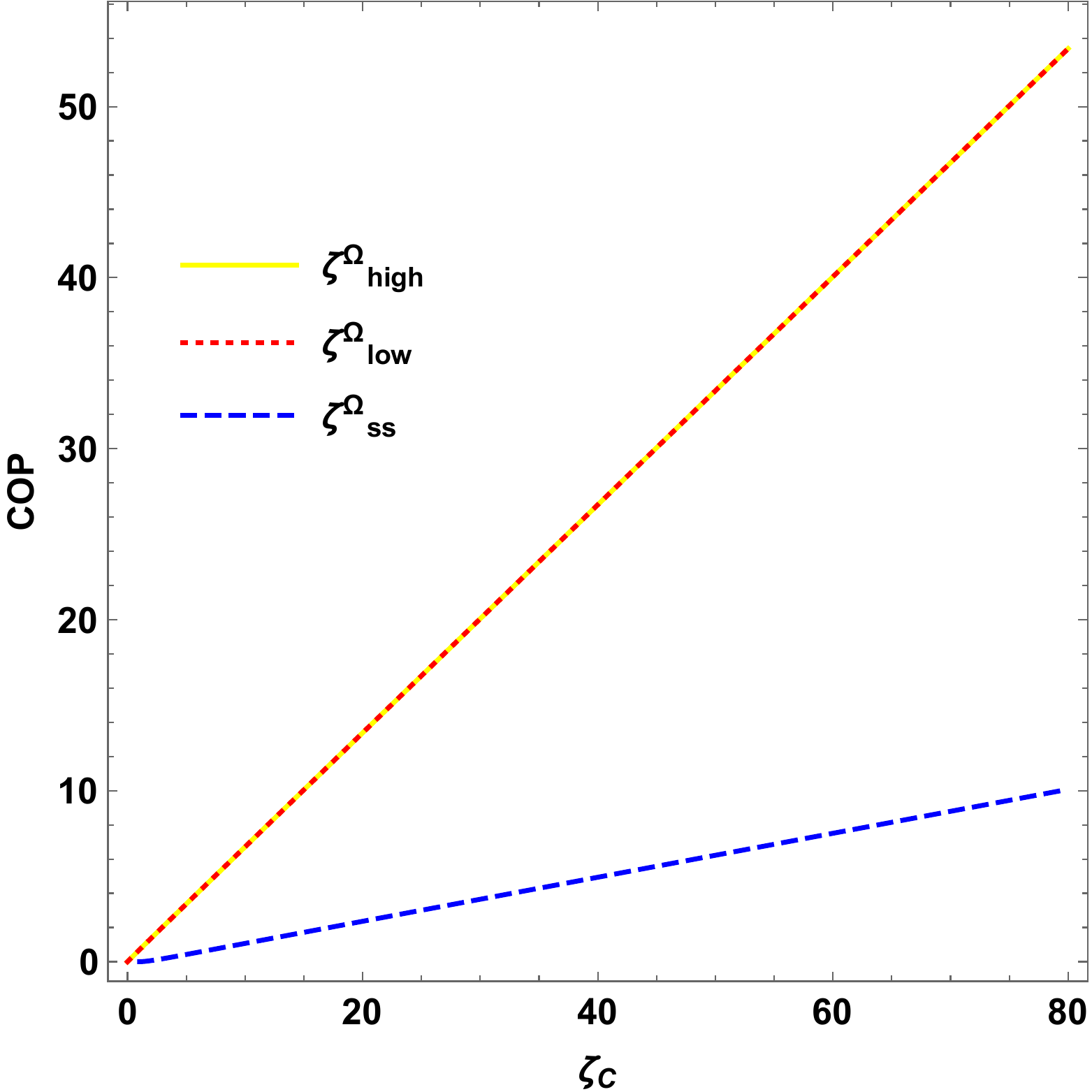}
 \end{center}
 \caption{COP at maximum $\Omega$ function and maximum $\chi$-function versus Carnot COP for Various cases discussed in the main text. All the notations used here are defined in the main text.} \label{copsgraph} 
\end{figure}
Using Eqs.  (\ref{coolingSS}) and (\ref{copmax}) in Eq. (\ref{omegaref1}), the desired form of $\Omega$ function can be found and optimized to yield the following optimal solution
\begin{equation}
z^* = \frac{ \zeta_{\rm C}\left[1+3 \zeta_{\rm C}-2\sqrt{2 \zeta_{\rm C}(1+ \zeta_{\rm C})}\right]}
{(1+ \zeta_{\rm C})\left[3(1+ \zeta_{\rm C})-2\sqrt{2 \zeta_{\rm C}(1+ \zeta_{\rm C})}\right]} \label{optimalz}.
\end{equation}
Using Eq. (\ref{optimalz}) in Eq. (\ref{copss}), we find the expression for the COP at MOF
as follows:
\begin{equation}
\zeta^{\Omega}_{ss} = \frac{A \left[1- \zeta_{\rm C}+A(1+ \zeta_{\rm C})\right]}{(A-1)\left[  \zeta_{\rm C} -A(1+ \zeta_{\rm C})  \right]}, \label{copssomega}
\end{equation} 
where $A=( \zeta_{\rm C}[1+3 \zeta_{\rm C}-2\sqrt{2 \zeta_{\rm C}(1+ \zeta_{\rm C})}]/(1+ \zeta_{\rm C})[3(1+ \zeta_{\rm C})-2\sqrt{2 \zeta_{\rm C}(1+ \zeta_{\rm C})}])^{1/2}$. We have plotted Eq. (\ref{copssomega}) in Fig. \ref{copsgraph} (dashed blue curve). As expected, COP for the sudden switch case is much smaller than the corresponding COPs obtained for the adiabatic case.

\subsection{Cooling power at maximum $\Omega$ function}
\begin{figure}
 \begin{center}
 \includegraphics[width=8.6cm]{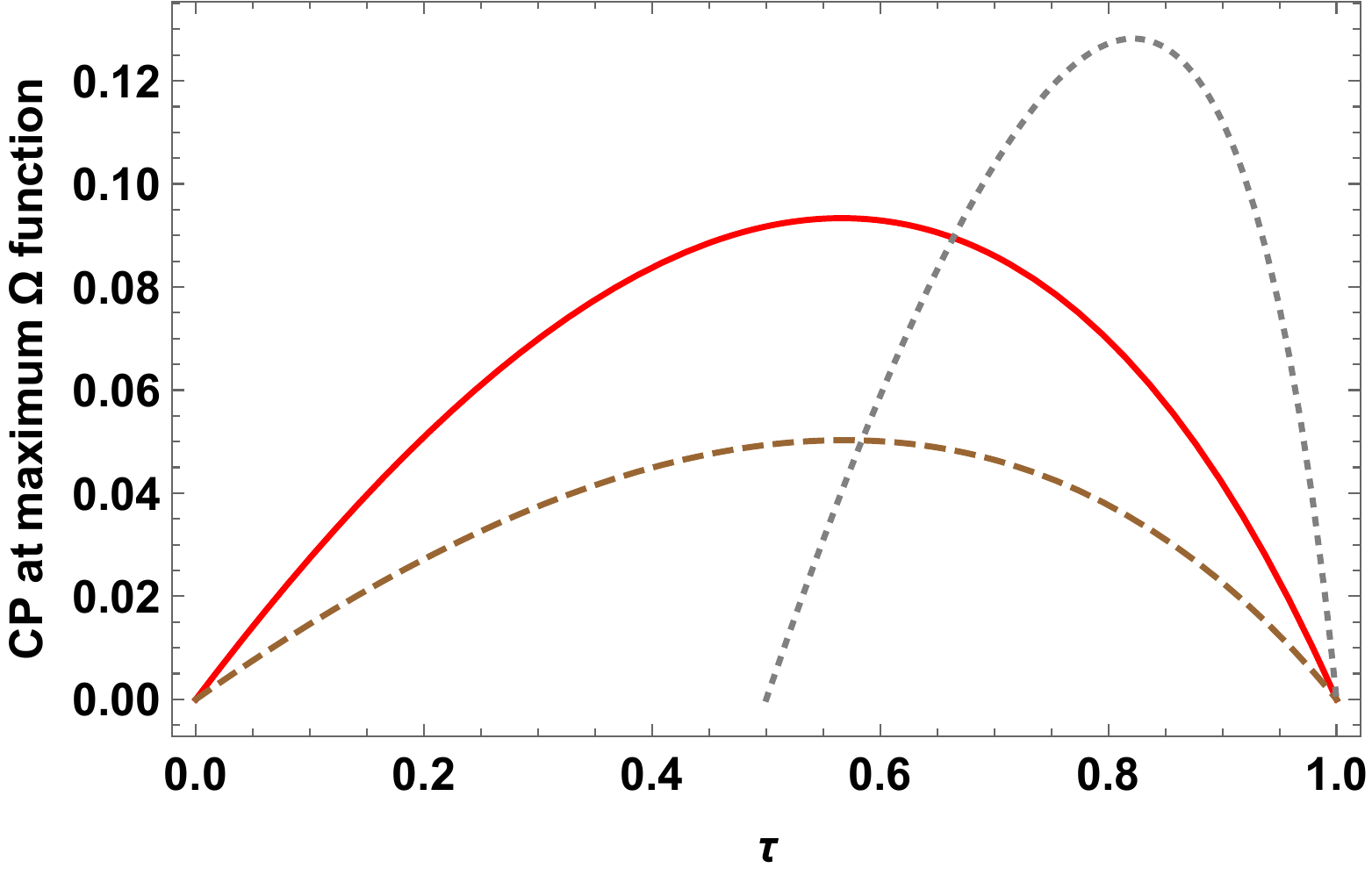}
 \end{center}
 \caption{COP at maximum $\Omega$ function is plotted against ratio of reservoir temperatures ($\tau$) for a fixed value of $\beta_2$, $\beta_2=1$. Solid red, dashed brown and dotted gray curves represent Eqs. (\ref{Qc1}), (\ref{Qc2}) and (\ref{Qc3}), respectively.} 
\end{figure}
In all the cases discussed above, cooling power is maximum at $z=0$, which is not a useful result. To look more into the behavior of cooling power, here, we will discuss the behavior of the cooling power at MOF. We start with the adiabatic case in the high temperature limit. In this case, substitution of the optimal solution $z^*=\tau/\sqrt{2-\tau}$ [see Eq. (\ref{zadoptimal})] in $Q_c=(\tau-z)/\beta_2$ yields the following expression for the CP
\begin{equation}
Q^{\Omega (\rm high)}_{4(\rm ad)} = \frac{1}{\beta_2}\left(\tau - \frac{\tau}{\sqrt{2-\tau}} \right). \label{Qc1}
\end{equation}
Similarly in the low-temperature regime for the adiabatic driving, substituting Eq. (\ref{coolLTsol}) in Eq. (\ref{coolLT}), we obtain
\begin{equation}
Q^{\Omega (\rm low)}_{4(\rm ad)} =  \frac{\left(\frac{1}{2-\tau }\right)^{\frac{1}{1-\tau }} \left[1-\tau -\ln \left(\frac{1}{2-\tau }\right)\right]\tau}{e \beta_2 (2-\tau )}. \label{Qc2}
\end{equation}
Finally, from Eqs. (\ref{optimalz}) and (\ref{coolingSS}), the CP at maximum $\Omega$-function for the sudden switch case is given by,
\begin{equation}
Q^{\Omega}_{4(\rm ss)} =  \frac{1}{2\beta_2} \left(2 \tau -\sqrt{\frac{\tau  \left(2 \tau -2 \sqrt{2 \tau }+1\right)}{3-2 \sqrt{2 \tau }}}-1\right). \label{Qc3}
\end{equation}
We plot Eqs. (\ref{Qc1}), (\ref{Qc2}) and (\ref{Qc3}) as a function of $\tau$ for a fixed value of $\beta_2$ in Fig. 6. It is clear from the figure that the maximum of CP exists at some value of $\tau$ for each case discussed above. This suggests that when we are operating the refrigerator at MOF, temperatures of the reservoirs can always be chosen so that they correspond to the maximum CP achievable. In this way, we can choose the optimal operating point for the refrigerator under consideration.  This kind of behavior is not exclusive to the harmonic Otto refrigerator. The same trend for the CP was observed for a three-level quantum refrigerator operating at MOF \cite{Kiran2021}.
\section{Conclusions}
We have investigated the optimal performance of a harmonic quantum Otto cycle working under the conditions of maximum $\Omega$ function. First, we obtained the analytic expressions for the EMOF of the engine in the adiabatic driving regime for high- and low-temperature regimes. In particular, for the engine in the low-temperature regime, we showed that the harmonic Otto engine can be mapped to a classical heat engine known as Feynman's ratchet and pawl model. Then, in the nonadiabatic driving regime in which we suddenly modulate the frequency of the oscillator from one value to another, we obtained loop-shaped curves for the efficiency-work plot characterizing the irreversible behavior of the engine under consideration. We repeated our analysis to study the optimal performance of the Otto refrigerator and obtained corresponding analytic expressions for the COP of the refrigerator.  Further, we explored the behavior of the cooling 
power under the conditions of MOF. 

\section{Acknowledgements}
This research was supported by the Institute for Basic Science in Korea (IBS-R024-Y2 and IBS-R024-D1). OA acknowledges support from the UK EPSRC EP/S02994X/1.
 
\bibliography{QHE-reference}
\bibliographystyle{apsrev4-2}

\end{document}